\documentclass[preprint]{aastex}
 
\shorttitle{Educational Spectrograph}
\shortauthors{Kannappan, Fabricant, \& Hughes}
 
\begin{document}
 
\title{Building a CCD Spectrograph for Educational or Amateur Astronomy}

\author{Sheila J. Kannappan\altaffilmark{1,2},
Daniel G. Fabricant\altaffilmark{1}, and Charles B. Hughes\altaffilmark{1}}

\altaffiltext{1}{Harvard-Smithsonian Center for
Astrophysics, 60 Garden St. MS-20, Cambridge, MA 02138;
skannappan@cfa.harvard.edu, dfabricant@cfa.harvard.edu, cn0412@hotmail.com}

\altaffiltext{2}{Current
address: The University of Texas at Austin, McDonald Observatory, 2511
Speedway, RLM 15.308 C1400, Austin, TX 78712; sheila@astro.as.utexas.edu}

\begin{abstract}

We discuss the design of an inexpensive, high-throughput CCD spectrograph
for a small telescope.  By using optical fibers to carry the light from the
telescope focus to a table-top spectrograph, one can minimize the weight
carried by the telescope and simplify the spectrograph design.  We recently
employed this approach in the construction of IntroSpec, an instrument
built for the 16-inch Knowles Telescope on the Harvard College campus.

\end{abstract}


\section{Introduction}

The widespread availability of affordable, high-quality commercial CCD
cameras for astronomy now makes it possible for educators and amateur
astronomers to conduct sophisticated observing programs with small
telescopes and limited resources.  In recent years the potential for
building inexpensive CCD spectrographs around these cameras has been
recognized by several companies and individuals
\citep[e.g.][]{glumac.sivo:building}.  Observations of the spectra of
stars, planets, and nebulae open up an exciting new realm of inquiry
for beginning astronomers, and the design and construction of the
spectrograph itself can provide students with a hands-on education in
the principles of optics and engineering.  Here we describe one
approach to designing and building a simple CCD spectrograph, based on
IntroSpec, an educational instrument we recently built for the
Harvard Astronomy Department.

We had four design goals for IntroSpec: (1) sufficient spectral range
and resolution for a variety of projects in optical astronomy, (2)
high throughput to enable observations of nearby nebulae and variable
stars, (3) simple controls and a durable design suitable for student
operation, and (4) a small telescope-borne weight.  We decided upon a
spectral range of $\sim$4500--6800~\AA\ at $\sim$6~\AA\ resolution.
This wavelength range includes many important spectral features: H,
Fe, Mg and Na absorption lines (prominent in stellar atmospheres),
molecular absorption bands from CH$_4$ and NH$_3$ (visible in
planetary atmospheres), as well as H, N, and O emission lines
(characteristic of star-forming nebulae).  Spectral coverage extending
$\sim$1000~\AA\ futher to the blue would be useful, but most
inexpensive CCD chips have poor UV response.  Our chosen resolution of
6~\AA\ resolution allows detection of most of the commonly observed
features in this spectral range and is adequate to resolve the closely
spaced emission lines of H$\alpha$ and [NII].

We were able to meet our design goals on a budget of slightly under \$5500,
of which $\sim$\$4000 went into buying a mid-range Meade CCD camera and
paying for the time and materials of a skilled machinist
(CBH).\footnote{These numbers do not include an expensive but desirable
accessory: a Hg-Ne lamp for wavelength calibration.  We purchased one from
Oriel for an additional \$450.}  Below we describe how IntroSpec's design
was tailored to the specific challenges posed by our budget and performance
requirements.  \S\ref{sc:optimizing} reviews the general optical design
constraints on any similar spectrograph, while \S\ref{sc:introspec}
examines the specific choices we made for IntroSpec, with special attention
to the instrument's fiber-fed design.

\section{Optical Design}
\label{sc:optimizing}

We begin with a brief review of the layout of a generic spectrograph,
then consider how to optimize its optical design. 

\subsection{Layout of a Generic Spectrograph}
\label{sc:genericlayout}

\begin{figure}[bt]
\epsscale{.5}
\plotone{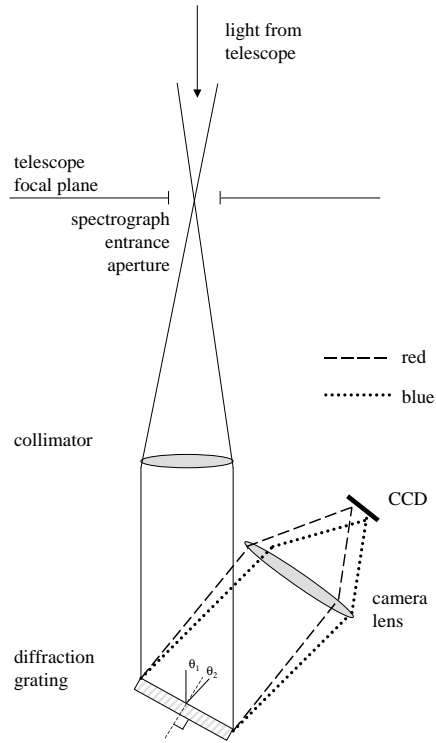}
\caption{Schematic of the optics in a simple spectrograph with a reflection
grating.  Because the angle between the grating normal and the camera's
optical axis ($\theta_2$) is smaller than the angle between the grating
normal and the collimator's optical axis ($\theta_1$), the diameter of the
monochromatic collimated beam leaving the grating has a larger diameter
than than the incident beam (anamorphic magnification).}
\label{fg:generic}
\end{figure}

Figure~\ref{fg:generic} shows the light path in a simple
CCD spectrograph.  The aperture accepts light from a small portion of the
telescope focal plane that may contain the image of a star or a portion of
an extended nebula. This light is collimated with a lens, producing a
parallel beam.  The beam travels to a diffraction grating, which disperses
the light into a spectrum described by the grating equation
\citep[e.g.][]{schroeder:astronomical}:
\begin{equation}
m\lambda = d(\sin{\theta_1} + \sin{\theta_2}),
\end{equation}
where $m$ is the diffraction order, $\lambda$ is the wavelength of the
light, $d$ is the grating line spacing, and $\theta_1$ and $\theta_2$
are the incoming and outgoing angles of the light with respect to the
normal to the grating surface (see Schroeder for sign conventions).

A reflection grating such as that shown in Figure~\ref{fg:generic}, folds
the light path as well as dispersing the light. In this case the angle
between the incident and diffracted beam typically must be $>$30$\degr$ to
allow the camera optics to stay clear of the collimated beam heading toward
the grating.  The grating normal generally points closer to the camera axis
than to the collimator axis \citep[][]{schroeder:astronomical}, leading to
anamorphic magnification: a geometrical expansion of the dispersed
collimated beam along the dispersion axis described by $AM
=\cos{\theta_2}/\cos{\theta_1}$ (see Figure~\ref{fg:generic}).  Anamorphic
magnification increases the spectral resolution by the same factor that the
collimated beam is expanded, but requires a larger camera acceptance
aperture to avoid vignetting.  Camera-collimator angles exceeding 45$\degr$
are rarely used for this reason.

The light dispersed by the grating is a superposition of collimated beams
of different colors, all traveling at slightly different angles.  The
camera lens focuses each color onto a location on the CCD chip determined
by its incoming angle, thereby forming a spectrum.  This focusing process
is exactly analogous to the way a telescope focuses incoming parallel light
beams from stars at different angles on the sky onto different positions on
the telescope focal plane: in the spectrograph, each point along the
spectrum is an image of the light from the spectrograph entrance aperture
at a different wavelength.  A higher dispersion grating increases the
angular spread between collimated beams of different colors, spreading the
corresponding images further apart.  For a detector of fixed size, higher
dispersion yields higher spectral resolution at the expense of reduced
spectral coverage.

\subsection{Setting the Relative Scales of the System}
\label{sc:sampling}

Independent of the grating choice or the absolute scale of the
spectrograph, we wish to specify the relative focal lengths of the
spectrograph optics so that the image of the spectrograph entrance aperture
is correctly sampled by the the CCD pixels.  The width of this image in the
dispersion direction defines the spectroscopic resolution element.
Oversampling the image wastes CCD pixels and decreases spectral coverage,
while undersampling it degrades spectral resolution. In the Nyquist
sampling limit, $\sim$2 pixels per wavelength allow one to recover the peak
and valley of an oscillating waveform.  Typically, finer sampling up to
3--4 pixels per resolution element is advantageous in professional
instruments with large CCD chips.  However, for an economical instrument,
CCD pixels for spectral coverage are at a premium, and one may prefer
sampling near the Nyquist limit.  With Nyquist sampling, IntroSpec's design
target of 6~\AA\ resolution combined with 2300~\AA\ spectral coverage
requires a CCD chip with $2\times2300/6=767$ pixels. We chose this
combination of spectral resolution and spectral coverage knowing that the
largest CCD within our budget had 768 pixels in the long dimension.

Once the desired sampling is chosen, the ratio of the focal lengths of the
collimator and camera lenses, $f_{\rm coll}$ and $f_{\rm cam}$, is
determined by the diameter ratio of the spectrograph entrance aperture and
its image.  For Nyquist sampling, the image of the spectrograph entrance
aperture must be reduced to $\sim$2 CCD pixels, or $\sim$20 $\mu$m for a
typical amateur CCD camera.  The spectrograph entrance aperture has
generally been sized to the expected diameter of a stellar image at the
telescope's focus, which is determined by atmospheric conditions and the
telescope optics. This choice maximizes the constrast between the the
object and the background sky light.  For the 40 cm $f$/10 Knowles
Telescope at Harvard, the local five-arcsecond seeing disk makes a 100
$\mu$m image at the telescope's focus, implying that $f_{\rm coll}/f_{\rm
cam}$ must equal $\sim$5 to reduce the image to $\sim$20 $\mu$m.  For a 20
cm $f$/10 telescope, similar seeing would make a 50 $\mu$m point source
image, requiring an $f_{\rm coll}/f_{\rm cam}$ ratio of $\sim$2.5.
However, a larger entrance aperture (implying higher $f_{\rm coll}/f_{\rm
cam}$ to achieve the same resolution) might be desirable for a 20 cm
telescope to allow for image shifts due to imperfect guiding.

The acceptance angles of the collimator and camera lenses can also be
estimated approximately independently of the absolute scale of the
spectrograph.  The collimator must be fast enough to accept an $f$/10 light
cone from the telescope from every point within the spectrograph entrance
aperture.  This aperture typically takes the form of a long slit to enable
spatial sampling perpendicular to the dispersion direction.  The camera
must be faster than the collimator: if $f_{\rm coll}/f_{\rm cam}\sim5$,
each monochromatic beam intercepted by the camera has a focal ratio of
$\sim$$f/2$ and is elongated in the dispersion direction by anamorphic
magnification (see \S\ref{sc:genericlayout}).  These beams must be accepted
over a field of view that is extended in both the spatial direction (due to
the slit length) and the dispersion direction (due to the angular
divergence of the different wavelengths in the spectrum,
\S\ref{sc:vignetting}).

\begin{deluxetable}{ccccccc}
\tablenum{1}
\tabletypesize{\footnotesize}
\tablewidth{0pt}
\tablecaption{Choosing the Camera Lens and Diffraction Grating}

\tablehead{

\colhead{Camera} &
\colhead{Collimator} &
\colhead{Collimated\tablenotemark{a}} & 
\colhead{Grating Ruling\tablenotemark{b}} &
\colhead{Spectral\tablenotemark{c}} &
\colhead{2 pixel} &
\colhead{Anamorphic} \\

\colhead{Foc.~Length} &
\colhead{Foc.~Length} &
\colhead{Beam Dia.} & 
\colhead{Density} &
\colhead{Range} &
\colhead{Spectral Res.} &
\colhead{Magnification} \\

\colhead{(mm)} &
\colhead{(mm)} &
\colhead{(mm)} &
\colhead{(lines/mm)} &
\colhead{(\AA)} &
\colhead{(\AA)} &
\colhead{}}

\startdata
35  &  175 & 17.5 & 900 &  2161 &  5.6 & 1.20 \\
50  &  250 & 25.0 & 600 &  2267 &  5.9 & 1.13 \\
85  &  425 & 42.5 & 300 &  2636 &  6.9 & 1.06 \\
\enddata 

\tablenotetext{a}{The clear apertures of the optical elements must
be larger than the nominal collimated beam diameter for the reasons
discussed in \S\ref{sc:sampling}.}
\tablenotetext{b}{Commonly available grating ruling closest to the ideal
grating ruling for the indicated $f_{\rm cam}$.  The ideal value may be
estimated to within $\sim$5\% from $\Delta\theta / \Delta\lambda_{spec}$,
where $\Delta\theta$ is the angular diameter of the CCD chip at $f_{\rm
cam}$ (in radians) and $\Delta\lambda_{spec}$ is the preferred wavelength
resolution (in mm).  We assume the grating is used in first order.}
\tablenotetext{c}{Spectral range calculated by determining the wavelength
falling at each end of the CCD chip, using the grating equation.}

\end{deluxetable}

\subsection{Setting the Absolute Scale of the System}
\label{sc:absolute}

We have some flexibility in choosing the overall dimensions of the optics,
limited by financial constraints and the desired spectroscopic performance.
We must balance two considerations.  First, larger optics and their mounts
are generally more expensive, although the economies of mass-produced
commercial optics can bend this rule.  Second, to maintain a given spectral
range and resolution on a fixed CCD format, one must keep the product of
the grating dispersion (ruling density) and the focal length of the camera
constant. The grating cost goes up nonlinearly with the ruling density, so
the grating cost and availability must be balanced against the cost of the
optics, particularly the camera lens.

The widespread availability of commercial 35 mm format single lens reflex
(SLR) photographic lenses suggests an attractive solution.  These lenses
are inexpensive, offer excellent image quality, have sophisticated
anti-reflection coatings, and are designed to work over a film format of 24
$\times$ 36 mm.  CCD chips in typical amateur cameras fit easily within
this format.  Many photographic lenses can produce images of $\sim$20
$\mu$m diameter, or about 2 pixels for a CCD camera with $\sim$10 $\mu$m
pixels. Excellent used lenses can be obtained for less than \$100,
particularly if an older manual focus lens is selected.

To minimize vignetting, our design for IntroSpec calls for a camera lens
faster than $f$/2 (\S\ref{sc:sampling}), so an $f$/1.4 lens is a good
choice.  Commercial 35 mm SLR lenses as fast as $f$/1.4 are typically
available with focal lengths between 35 mm and 85 mm, with 50 mm being by
far the most common.  These focal lengths pair conveniently with readily
available diffraction gratings to meet our goals for spectral coverage and
resolution.  Table~1 describes three combinations of gratings and lenses
with focal lengths between 35 and 85 mm.  We assume a
camera-collimator angle of 36$^{\circ}$, a central wavelength of 5650~\AA,
and a 6.9 mm CCD chip with 768 pixels (as found in IntroSpec).  All of
these lens-grating combinations approximately satisfy our design targets of
6~\AA\ resolution and 2300~\AA\ spectral coverage.  Given the ease of
obtaining both 50 mm focal length lenses and 600 line/mm gratings, we
selected this configuration for IntroSpec.  The spectrograph would have
been somewhat smaller had we used a more expensive 35 mm focal length
camera lens, but as Table~1 demonstrates, vignetting due to anamorphic
magnification would be slightly greater.

\section{IntroSpec}
\label{sc:introspec}

\subsection{Optical Fibers Ease Design Constraints}

IntroSpec's design follows that of the generic CCD spectrograph in
Figure~\ref{fg:generic}, with one key difference: the spectrograph
does not mount directly to the telescope.  This change reflects the
optical design constraints outlined in \S\ref{sc:optimizing}, which
led us to prefer moderately large optical elements (a 50 mm focal
length camera lens and a 250 mm focal length collimator), implying an
instrument 
\clearpage

\noindent whose longest dimension exceeds 0.5 meter.  A commercial 40
cm telescope can support neither the size nor the weight of such an
instrument.  Instead of sacrificing optical performance, we opted for
a bench-mounted, fiber-fed design.  Light from the telescope enters a
set of optical fibers at the telescope's focus, and from there the
fibers carry the light to IntroSpec, which sits on a table.  Because
IntroSpec's weight is removed from the telescope, it can be packaged
in a heavy metal box with an easy-to-remove lid, which protects the
instrument while keeping it accessible to students.  IntroSpec's
wavelength calibration is stable because the mechanical flexure
encountered in a telescope-mounted spectrograph is eliminated.

\subsection{General Overview of the Instrument}
\label{sc:overview}

\begin{figure}[bt]
\epsscale{1.}
\plotone{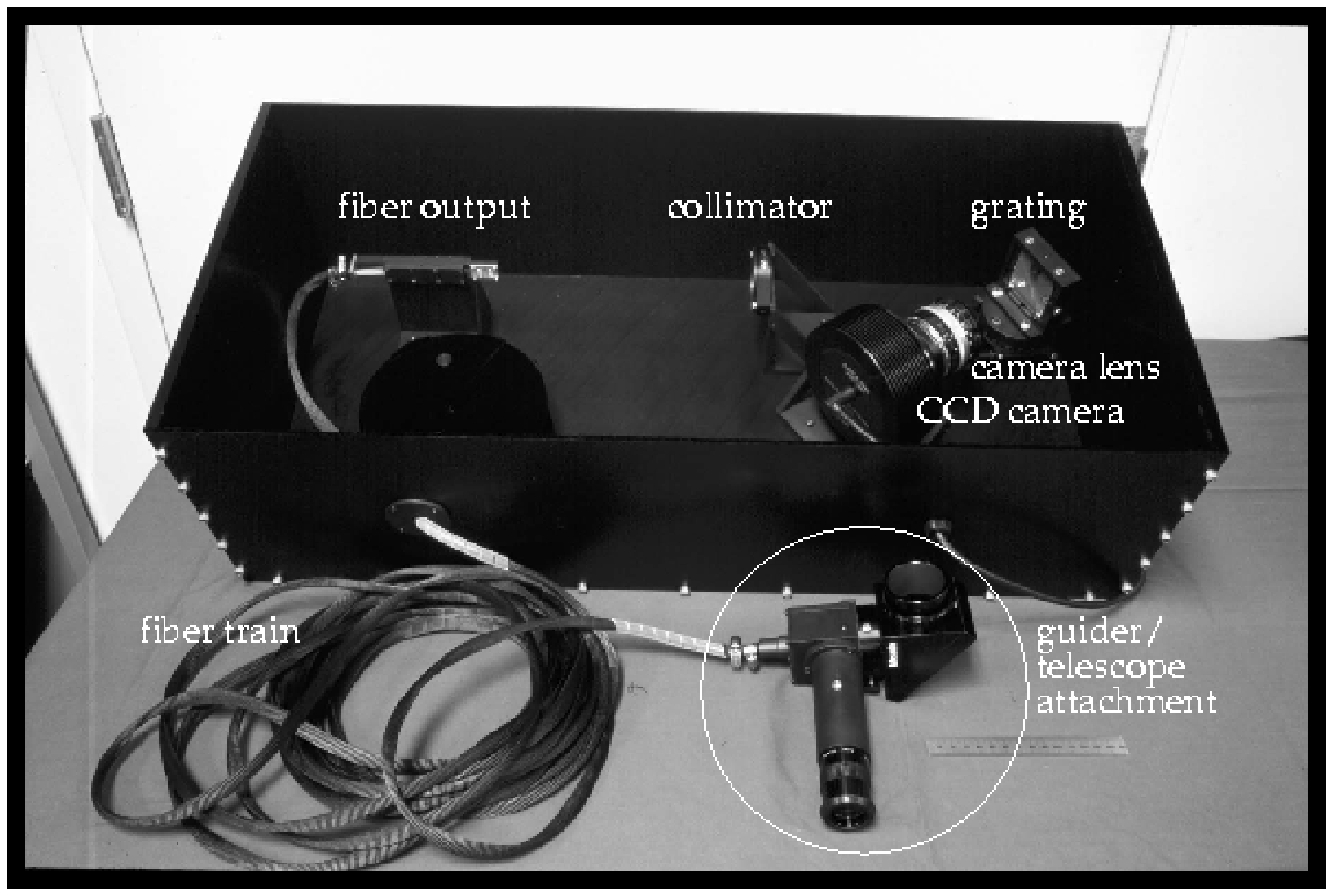}
\caption{Overview of IntroSpec.  See \S\ref{sc:overview} and
Table~2.}
\label{fg:introspecphoto}
\end{figure}

\begin{figure}[bt]
\epsscale{0.65}
\plotone{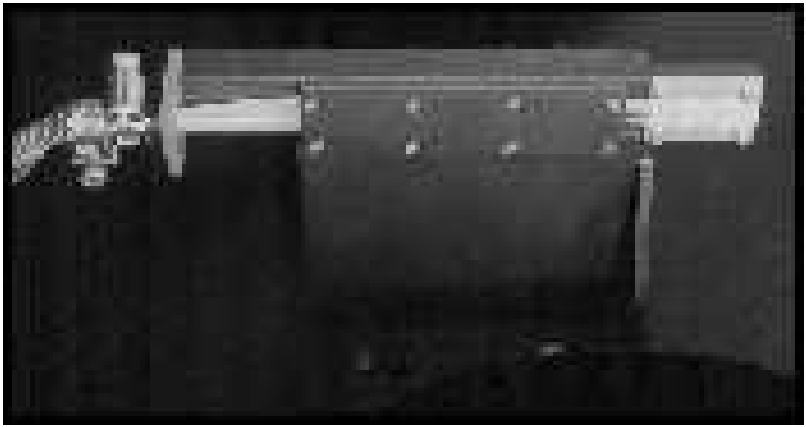}
\caption{Fiber output assembly.  Three pairs of fibers with diameters
of 100, 200, and 300 $\mu$m are arranged vertically, with the smallest
fibers at the center and the largest fibers at the top and bottom. The
ends of the fibers are aligned precisely in six V-grooves; the
V-grooves are bead blasted but not anodized (see
\S\ref{sc:working}).}
\label{fg:fiberoutput}
\end{figure}

\begin{figure}[bt]
\epsscale{.65}
\plotone{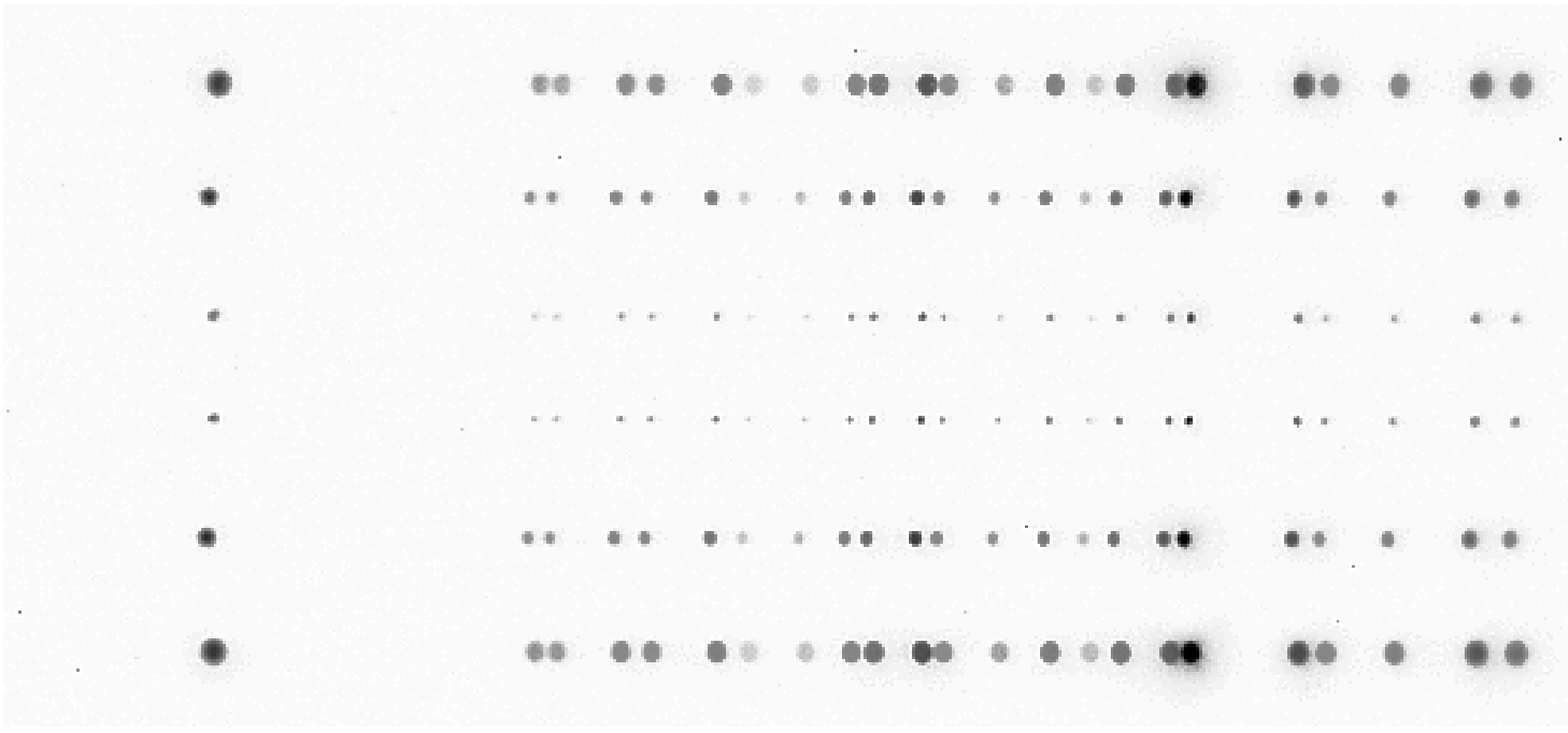}
\caption{CCD image of spectra from IntroSpec's six
optical fibers.  The fibers have been illuminated with a mercury-neon
calibration lamp, which emits a number of narrow emission lines.  Here
the camera lens was opened all the way to $f/1.4$ and a small amount
of light in a diffuse halo is visible surrounding the brighter lines.}
\label{fg:compimage}
\end{figure}

Figure~\ref{fg:introspecphoto} and Table~2 provide an overview of IntroSpec.
Optical fibers are coupled to the telescope via a guider/telescope
attachment, further discussed in \S\ref{sc:manual}.  The fiber train
consists of six fibers, each of which accepts light from a different
position on the telescope focal plane; these are bundled in several
layers of protective coverings.  (\S\ref{sc:working} discusses the
technical aspects of working with fibers in more detail.)  Upon
entering the spectrograph enclosure, the fiber train loops around a
large black spool for strain relief, then the fibers are channeled
into the fiber output, from which the bare fiber ends emerge in a
vertical stack (Figure~\ref{fg:fiberoutput}).  The base of the fiber
output assembly is adjustable to permit focusing the collimator lens. 
Each optical fiber is an independent
spectrograph entrance aperture; by stacking these apertures
perpendicular to the dispersion axis, we obtain six parallel spectra
on the CCD chip (Figure~\ref{fg:compimage}).

\begin{deluxetable}{l}
\tablenum{2}
\tabletypesize{\footnotesize}
\tablewidth{0pt}
\tablecaption{Purchased Parts List\tablenotemark{a}}
\tablehead{\colhead{}}

\startdata
\cutinhead{Spectrograph}
Newport PAC088 cemented achromat, 2" diameter, $f = 250 $ mm \\
Newport LH-200 2" diameter lens mount \\
Richardson Grating Lab 588 lines/mm reflection grating,  \\
\phm{make a big space}$50\times50\times6 $ mm blazed at 561 nm  \\
Newport 481-A Rotation Stage  \\
Nikon $f/1.4 $ $f = 50 $ mm photographic lens \\
Meade Pictor 416 XT CCD camera \\
Meade 12 V DC to 115 AC adapter \\
1.25" CCD barrel from Meade Customer Service \\
surplus bayonet mount (free from photography shop) \\

\cutinhead{Guider}
2 Newport PAC040 precision achromatic doublets, 25.4 mm diam, EFL = 50.8 mm \\
Orion Sirius Plossl eyepiece 1.25" diameter, $f = 26 $ mm \\
Meade 2" diagonal mirror \#929  \\

\cutinhead{Fiber Train}
Polymicro FVP series multimode step-index fiber, core diam 100 \& 300 $\mu$m \\
Fiberguide SFS200 240T multimode step-index fiber, core diam 200 $\mu$m \\
M M Newman 20 gauge PTFE standard wall Teflon tubing \\
Bally Ribbon Mills Pattern 8444 0.5'' Tubular Nylon Webbing \\
\enddata

\tablenotetext{a}{All other parts were made in the machine shop or the lab.}
\end{deluxetable}

IntroSpec uses two fibers of each of three diameters: 100, 200, and
300 $\mu$m.  The seeing disk at the Knowles telescope is well matched
to a 100 $\mu$m fiber, and this fiber diameter provides the highest
spectral resolution, $\sim$6~\AA\ FWHM.  Larger fibers give the user
the choice to trade off spectral resolution in exchange for gathering
additional light from an extended source such as a nebula. Larger
fibers are also more forgiving of guiding errors when spectral
resolution is not critical.  The different fiber diameters allow
students to learn about the instrument and to make the connection
between fiber size and spectral resolution.  For point sources, one
fiber can be used to acquire the target spectrum and the second fiber
of the same diameter can be used to acquire a sky spectrum for
background subtraction.

From the collimator on, IntroSpec looks much like the generic
spectrograph of Figure~\ref{fg:generic}.  The collimator lens is a 50
mm diameter achromat with $f_{\rm coll}=250$ mm, yielding a collimated
beam diameter of 25 mm with an f/10 input light cone.  We selected a
588 line mm$^{-1}$ grating blazed\footnote{The ruled diffraction
gratings typically used in astronomical instruments are blazed to
maximize the amount of light in the desired diffraction order.  A
blazed reflection grating has a triangular groove profile, with the
reflective surface of the groove inclined to the grating surface by an
angle known as the blaze angle.  The blaze angle is chosen such that
simple geometrical reflection from the groove face sends light not
into the central maximum, but instead into the chosen diffraction
order, coincident with the desired peak-efficiency wavelength. In the
case of transmission gratings, the blaze angle is chosen such that
light geometrically refracted from the groove face is coincident with
the desired peak-efficiency wavelength in the selected diffraction
order. We chose a reflection grating for IntroSpec because reflection
gratings are somewhat less expensive than transmission gratings and
the folding of the light path is convenient.} for peak efficiency at
5610~\AA.  Richardson Grating Laboratory offers this grating in sizes
up to $\sim$50 mm square.  The grating face is precisely aligned to
the rotation axis of a rotation stage, permitting micrometer
adjustment of the grating angle to vary the spectral range.  With a
camera-collimator angle of 36$\degr$, IntroSpec's design range of
4500--6800~\AA\ corresponds to an 8$\degr$ angle between the grating
normal and the camera axis.  At this rotation angle, the anamorphic
magnification factor is 1.1, so the monochromatic outgoing beams are
oval in shape with dimensions $\sim$25 $\times$ 28 mm.

These beams travel $\sim$90 mm from the grating to the camera lens,
where the dispersed colors slightly overfill the 38-mm maximum
aperture of the camera lens, a 50 mm focal length $f/1.4$ Nikon lens
(see the discussion of vignetting in \S\ref{sc:vignetting}).  The lens
focuses the spectrum onto a Meade Pictor 416 XT CCD camera, containing
a $768\times512$ KAF-0400 chip with 9 $\mu$m square pixels. Attaching
the Nikon lens to the Pictor 416 XT required machining a special
connector made from a Nikon bayonet mount and a Meade CCD camera
mount.  Finally, the cord shown exiting the spectrograph enclosure to
the front right in Figure~\ref{fg:introspecphoto} connects the CCD
camera to the Meade Pictor control box (not shown) for data and power
transfer.

\subsection{Image Quality \& Vignetting}
\label{sc:vignetting}

We have modeled IntroSpec's optical performance using the commercially
available optical design and analysis code ZEMAX.  The optical prescription
for the collimator achromat is included in the lens catalog supplied with
ZEMAX.  The precise optical prescription for the Nikon lens is not
available, so for this lens we have substituted the scaled prescription for
a 100 mm focal length, $f/1.4$ Double Gauss lens from a compilation of lens
designs by \citet{smith}.  The Nikon lens used in IntroSpec is also of the
Double Gauss design and should perform similarly.

Our raytraces for the model IntroSpec give images with RMS image
diameters between 12 and 30 microns across the target 4500 to
6800~\AA\ wavelength range when the camera lens is stopped down to
$f/1.8$.  The performance is considerably worse when the lens is
opened to its full $f/1.4$ aperture.  Our laboratory tests show that
the actual Nikon lens produces CCD images with FWHM diameters of
$\sim$20--25 $\mu$m when the grating is illuminated with collimated
light, in good agreement with the raytraces.

\begin{figure}[bt]
\epsscale{.45}
\plotone{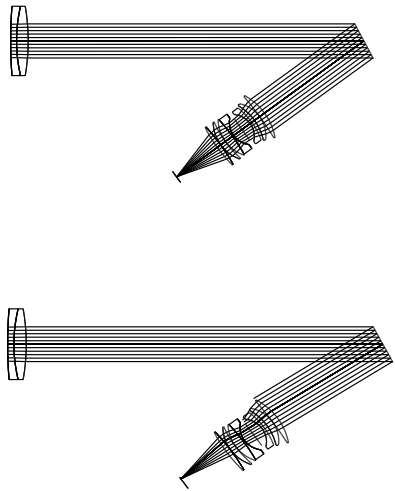}
\caption{The optical light path from the collimator to the focal
surface for two wavelengths, near 5650~\AA\ in the top panel and
near 6800~\AA\ in the lower panel.  The camera lens is stopped
down to f/1.8.  Approximately 32\% of the light is lost to vignetting
at 6800~\AA.}
\label{fg:zemax}
\end{figure}

Our raytraces also allow us to estimate the vignetting in the camera lens
as a function of wavelength. As shown in Figure~\ref{fg:zemax}, when the
camera lens is stopped down to $f/1.8$, the on-axis (5650~\AA) beam is
unvignetted, and $\sim$68\% of the extreme (4500~\AA\ or 6800~\AA) beams
are transmitted.

\clearpage

\subsection{Throughput \& Optical Fiber Losses}
\label{sc:working}

We estimate IntroSpec's total throughput at about 12\% at 5650~\AA.
Throughput is lower at the ends of the spectral range due to
vignetting in the camera lens, the decrease of the grating efficiency
off the blaze peak, and the wavelength-dependence of the quantum
efficiency of the CCD camera.

Table~3 lists the individual throughput of each of Introspec's optical
elements, not including losses due to guiding errors or the finite
apertures of the fibers.  With high-quality components, the lens and
grating efficiencies are hard to improve.  However, current versions
of the Meade Pictor CCD cameras use CCDs that have quantum
efficiencies of $\sim$50\% at 5650~\AA, noticeably higher than the
35\% quantum efficiency of the older model in IntroSpec.

\begin{deluxetable}{lc}
\tablenum{3}
\tabletypesize{\footnotesize}
\tablewidth{0pt}
\tablecaption{Estimated Introspec Throughput at 5650~\AA}

\tablehead{

\colhead{Optical} &
\colhead{Transmission} \\

\colhead{Element} &
\colhead{}
}

\startdata
Fibers		& 0.60\\
Collimator Lens	& 0.97\\
Grating		& 0.70\\
Camera Lens	& 0.86\\
CCD		& 0.35\\
Total Throughput& 0.12\\
\enddata 

\end{deluxetable}

Achieving the optical fiber throughput shown in Table~3 requires some
attention to minimizing fiber losses.  IntroSpec uses multimode, step-index
fibers with a pure fused silica core and doped fused silica cladding,
because this type of fiber provides the best overall throughput for the
wavelengths of interest in astronomical spectroscopy.

Fiber losses come from three main sources: end reflection losses, internal
transmission losses, and focal ratio degradation.  End reflection losses
are simply the Fresnel reflection losses encountered at every material
interface; for fused silica, the refractive index is $\sim$1.46 at the
center of the IntroSpec passband, resulting in a loss of $\sim$3.5\% at
each fiber end.  These losses are hard to avoid as it is usually not
practical to apply antireflection coatings to the tiny fiber ends.
Internal transmission losses can be estimated using data from the
fiber manufacturer and vary according to the OH content of the fused silica
and the material preparation.  For the 10 meter long fibers used in
IntroSpec, internal transmission losses are negligible in the 4500 to
6800~\AA\ range, although transmission losses become more important below
4000~\AA.  A discussion of the internal transmission of modern fused silica
fibers can be found in \citet{lu}.

Focal ratio degradation is the major controllable source of fiber
losses.  If a fiber is manufactured to high standards and not pinched
in use, a light cone incident on the fiber axis exits the fiber with
approximately the same focal ratio with which it entered.  (The fibers
we have chosen accept a numerical aperture of 0.22, corresponding
to a focal ratio of $\sim$$f/2.2$, much faster than the required
$f/10$.)  However, stresses that locally deform the fiber will scatter
light into a faster cone, resulting in focal ratio degradation.
This degradation in turn leads to reduced system throughput if the
optics are not enlarged to capture the faster cone of light.  Our
tests have shown that even if fibers are carefully mounted, protected
from stress, and properly aligned, focal ratio degradation will
scatter about 25\% of the light incident at $f/10$ outside
an $f/10$ light cone.  Accounting for reflection losses at both ends
of the fiber, the total fiber throughput under ideal conditions is
then about 70\% at $f/10$.  Focal ratio degradation is much less
important for input beams significantly faster than $f/10$.  We
recommend avoiding fibers with core diameters smaller than 100 $\mu$m
because they are more difficult to handle and are more prone to focal
ratio degradation.

\subsection{Mounting Optical Fibers}

The most common source of excess focal ratio degradation is external stress
due to inappropriate fiber mounting techniques.  For example, if the fibers
are bonded into an oversized ferrule with a circumferential epoxy bond, the
epoxy shrinkage upon curing can introduce serious stress.  A much better
technique is to bond the fibers onto precisely machined V-grooves (long
grooves with a V-shaped cross-sectional profile) with small beads of
silica-filled epoxy.  Mixing epoxy and silica half and half by weight
reduces the epoxy shrinkage and the viscous mixture does not completely
surround the fiber (Figure~\ref{fg:fiberoutput}).

To machine the V-grooves, we use a 45$\degr$ double angle cutter in an R-8
or 3/4'' straight shank stub milling machine arbor.  Cutters with
noticeable radii on the cutting teeth should be avoided, because a large
radius will disturb the fiber alignment by allowing the fiber to contact
the groove bottom.  After the grooves are cut, we lightly bead blast them
with glass beads; this process removes burrs that may stress or scratch the
fibers and also creates a good surface for bonding.

Stress along the length of the fiber can be minimized by enclosing each
fiber in a loosely fitting sleeve of teflon tubing (in addition to any
plastic buffer the fiber may already be coated with); for convenience and
extra protection, we also bundle these tubes within a larger fabric sheath.
Several stages of strain relief at both ends of the fiber train protect the
fibers from accidental tugs and maintain a large bend radius.

Accurate fiber alignment and polishing is also critical, both to avoid
focal ratio degradation losses at the entrance fiber face and
misdirection of the exiting light at the exit fiber face.  Both
the entrance and exit ends of the fibers must be polished flat to a
good optical finish, and the polished ends must be perpendicular to
the fiber axes (which are aligned perpendicular to the focal plane
using the precisely machined V-grooves).  We used a fiber polishing
machine built for an earlier project; a similar machine could be
constructed for $\sim$\$300 in parts.  The machine consists of a flat
glass disk attached to an aluminum turntable rotating on a 72 rpm
motor.  A rigid arm holds a brass cylinder perpendicular to the glass
disk.  The brass cylinder has V-grooves cut along its length, and the
fibers are temporarily attached to the V-grooves with Duco cement for
polishing.  A large drop of Duco cement covers the end of each fiber,
which protrudes by $\sim$2 mm (most of this will be polished away).
We attach disks of high quality self-adhesive abrasive paper in
successively finer grades to the turntable, and the brass cylinder
containing the fibers is swept across the abrasive paper.  The
sequence begins with 30 $\mu$m grit paper and ends with 0.3 $\mu$m
grit paper.  A thin solution of dishwashing soap in water provides
adequate lubrication; clean water is used to rinse off grit residue
before moving to the next smaller grit (the motor is isolated from
water spills).  Finally, we inspect the fiber ends under a microscope
to verify that no scratches remain.  The Duco cement can be removed
with acetone to release the fiber from the polishing cylinder.

\subsection{The Manual Guider}
\label{sc:manual}

Acquiring and tracking astronomical targets requires some technique for
viewing the focal plane.  In a professional instrument this function is
usually performed by a sensitive, high frame rate electronic camera. In
IntroSpec we have provided optics that allow the observer to visually
monitor and correct the telescope tracking while viewing a 12 arcminute
field of view imaged around the spectrograph entrance aperture.

\begin{figure}[bt]
\epsscale{.75}
\plotone{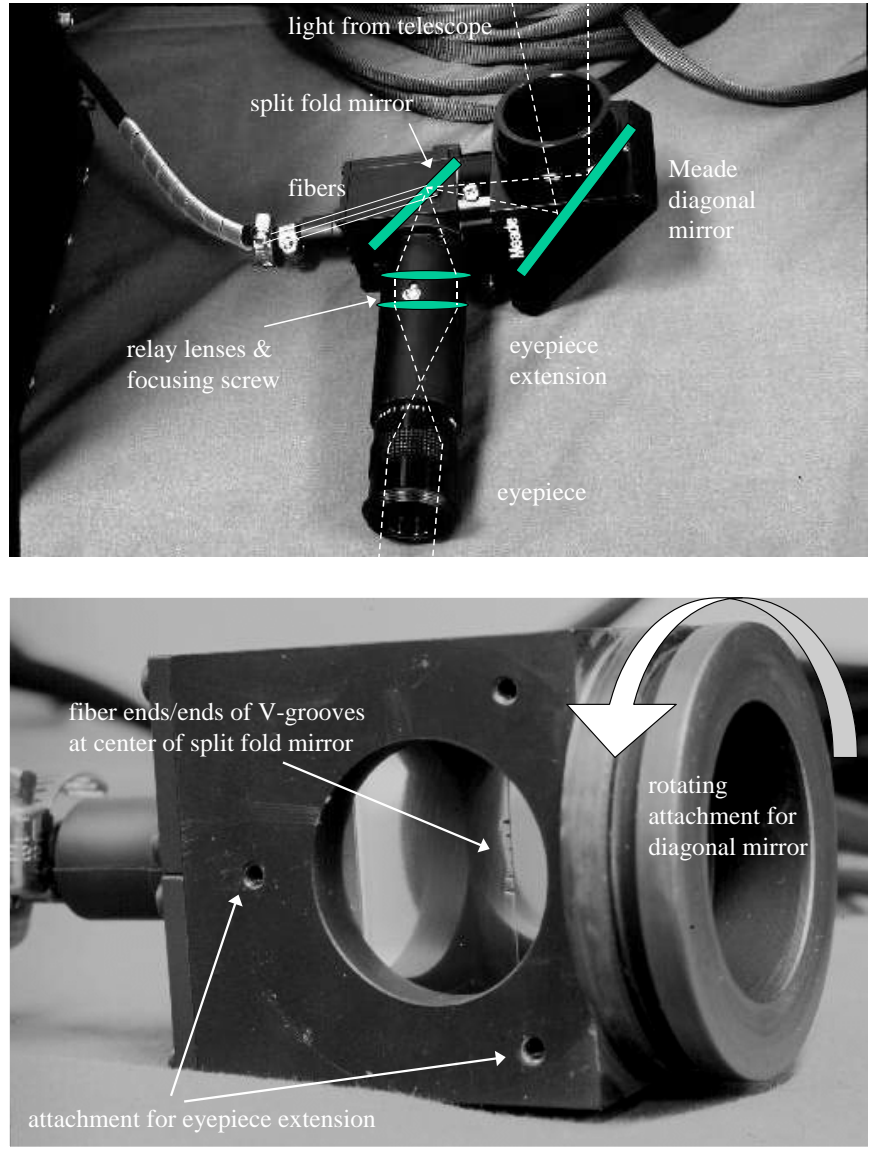}
\caption{The guider/telescope attachment.  {\em Upper Panel:} Lines
drawn over the photo show the positions of the optical components and
the path of the light.  The small screw between the relay lenses
allows them to slide forward or backward as a unit to focus. {\em
Lower Panel:} Inner guider assembly, with the Meade diagonal mirror
and the eyepiece extension removed to reveal the split mirror in the
central piece.  The fibers are not easily visible in the photo, but
they protrude slightly from the V-grooves, with their polished
surfaces flush to the focal plane of the telescope (i.e.\ at a
45$\degr$ angle from the mirror surface).}
\label{fg:guiderphotos}
\end{figure}

Figure~\ref{fg:guiderphotos} illustrates the primary components of
the guider.  A flat metal mirror intersects the focal plane of the
telescope at a 45$\degr$ angle. The mirror is machined in two pieces and
the fibers protrude through the mirror at the intersection of the focal
plane and the mirror's surface.  All the light that does not enter the
fibers strikes the mirror and is reflected by 90$\degr$ into the eyepiece
extension.  Inside the extension, two matched achromatic lenses relay the
image of the telescope focal plane to a 26 mm Plossl eyepiece.  The
observer guides the telescope by centering the object of interest on one of
the fiber ends so that most of the light disappears down the fiber.

The guider also serves as the mechanical interface between the telescope
and the fibers, incorporating mechanisms for eyepiece rotation, strain
relief, and fiber alignment.  To minimize weight, nearly all of the parts
were made from aluminum, including the mirror.  We finished the mirror in
two steps, first removing machining marks with a fine grit paste on a
turntable and then polishing with white rouge compound and a hard felt
wheel mounted on a lathe. The two halves of the mirror were registered with
pins, allowing us to disassemble the mirror to mount the fibers in internal
V-grooves without disturbing the polished surface.

\section{Conclusion}

IntroSpec saw first light on March 17, 1999 and is now in regular use in a
Harvard astronomy course for non-majors.  Students perform projects such as
comparing the spectra of the two stars

\begin{figure}[bt]
\epsscale{0.75}
\plotone{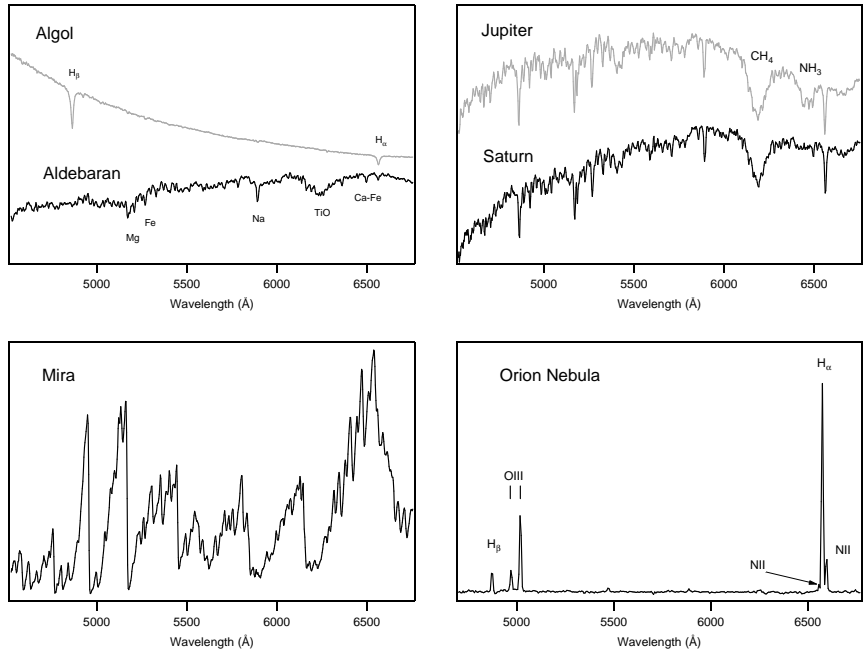}
\caption{Spectra taken with IntroSpec at the Knowles Telescope.
Approximate exposure times: Algol 30 sec; Aldebaran 3 sec; Jupiter 3
sec; Saturn 6 sec; Mira 30 sec; Orion Nebula 600 sec.  Observations
were taken on different nights with variable (but never photometric)
sky conditions.  All spectra except Orion were taken with the 100
$\mu$m fiber for maximum spectral resolution ($\sim$6~\AA); for Orion
we binned the CCD by two but still used the 100 $\mu$m fiber because
of readout glitches in the 200 $\mu$m fiber spectrum.  The resulting
spectral resolution of $\sim$12~\AA\ just suffices to distinguish the
NII line at 6548~\AA\ from H$\alpha$ at 6563~\AA.  The sawtooth
structure in Mira's spectrum is caused by absorption bands of TiO and
other molecules in its outer envelope.}
\label{fg:spectra}
\end{figure}

\clearpage

\noindent in Albireo.  Even under a bright
urban sky, we have successfully used IntroSpec to observe objects ranging
from stars and planets to emission nebulae and binary accretion systems.
Figure~\ref{fg:spectra} shows a few spectra
taken with IntroSpec along with their exposure times.  We have also used
IntroSpec to demonstrate to students how a spectrograph works; the
instrument's easy-to-remove top cover allows students to peer directly
inside.  In another article \citep{kannappan.fabricant:getting} we provide
a brief introduction to amateur spectroscopy with low-budget spectrographs
like IntroSpec, including basics of data analysis and interpretation as
well as a variety of spectroscopy project ideas.

\acknowledgements
IntroSpec was designed and built at the Harvard-Smithsonian Center for
Astrophysics, with financial support from Harvard College and the
Harvard Astronomy Department arranged by Jonathan Grindlay and Ramesh
Narayan.  We thank Professor Grindlay for his participation in
planning and commissioning the instrument.  The Knowles Telescope
where IntroSpec is used was generously donated by C. Harry and Janet
Knowles.  We are grateful to John Geary, Steve Amado, Joe Zajac,
Douglas Mar, John Roll, Warren Brown, and Anita Bowles for advice,
resources, and technical assistance during IntroSpec's construction
phase.  We also thank Tom Narita, John Raymond, Dimitar Sasselov, and
the students, friends, and family who helped with commissioning runs.

\end{document}